\documentclass[fleqn,usenatbib]{mnras}

\usepackage{newtxtext,newtxmath}

\usepackage[T1]{fontenc}
\usepackage{ae,aecompl}

\usepackage{graphicx}	
\usepackage{amsmath}	
\usepackage{amssymb}	
\usepackage[nolist]{acronym}
\usepackage{dsfont}
\usepackage{ulem}
\usepackage{breakcites}

\title[Representative turbulence profiles for ESO Paranal]{Representative optical turbulence profiles for ESO Paranal by hierarchical clustering}

\author[O. J. D. Farley et al.]{O. J. D. Farley,$^{1}$\thanks{E-mail: o.j.d.farley@durham.ac.uk}
J. Osborn,$^{1}$
T. Morris,$^{1}$
M. Sarazin,$^{2}$
T. Butterley,$^{1}$
M. J. Townson,$^{1}$
\newauthor P. Jia$^{3, 1}$
and R. W. Wilson$^{1}$
\\
$^{1}$Centre for Advanced Instrumentation (CfAI), Durham University, UK\\
$^{2}$European Southern Observatory, Karl-Schwarzshild-Str.2, 85748 Garching bei Muenchen, Germany\\
$^{3}$College of Physics and Optoelectronics, Taiyuan University of Technology, Taiyuan, 030024, China\\
}

\date{Accepted XXX. Received YYY; in original form ZZZ}

\pubyear{2018}

\begin{document}
\label{firstpage}
\pagerange{\pageref{firstpage}--\pageref{lastpage}}
\maketitle

\begin{acronym}
\acro{AO}{adaptive optics}
\acro{AT}{auxiliary telescope}
\acro{ELT}{extremely large telescope}
\acro{WFS}{wavefront sensor}
\acro{DM}{deformable mirror}
\acro{LGS}{laser guide star}
\end{acronym}

\begin{abstract}
Knowledge of the optical turbulence profile is important in \ac{AO} systems, particularly tomographic \ac{AO} systems such as those to be employed by the next generation of 40 m class \acp{ELT}. Site characterisation and monitoring campaigns have produced large quantities of turbulence profiling data for sites around the world. However \ac{AO} system design and performance characterisation is dependent on Monte-Carlo simulations that cannot make use of these large datasets due to long computation times. Here we address the question of how to reduce these large datasets into small sets of profiles that can feasibly be used in such Monte-Carlo simulations, whilst minimising the loss of information inherent in this effective compression of the data. We propose hierarchical clustering to partition the dataset according to the structure of the turbulence profiles and extract a single profile from each cluster. This method is applied to the Stereo-SCIDAR dataset from ESO Paranal containing over 10000 measurements of the turbulence profile from 83 nights. We present two methods of extracting turbulence profiles from the clusters, resulting in two sets of 18 profiles providing subtly different descriptions of the variability across the entire dataset. For generality we choose integrated parameters of the turbulence to measure the representativeness of our profiles and compare to others. Using these criterion we also show that such variability is difficult to capture with small sets of profiles associated with integrated turbulence parameters such as seeing.


\end{abstract}




\section{Introduction}
\acresetall
In tomographic \ac{AO}, multiple \acp{WFS} and \acp{DM} are used to measure and correct the turbulence in the Earth's atmosphere over a wide field of view. This wide corrected field has made tomographic \ac{AO} systems desirable for both current 8 m class telescopes \citep[see e.g.][]{Esposito2016,Neichel2014} and the next generation of 40 m class \acp{ELT} \citep[see e.g.][]{Diolaiti2010,Herriot2014,Hinz2010}.

In combining the offaxis \ac{WFS} measurements to reconstruct the  three dimensional volume of turbulence projected from the telescope pupil through the atmosphere, some knowledge of the vertical distribution of the turbulence is required \citep{Fusco2001,Vidal2010}. As such the performance of these systems depends on the optical turbulence profile, usually defined in terms of distribution of the refractive index structure constant $C_n^2(h)$ with altitude $h$. In particular, high altitude turbulence where the spatial overlap between \ac{WFS} measurements is small results in a degradation in \ac{AO} performance. 

The turbulence profile therefore plays a key role in the design of tomographic \ac{AO} systems as they must be optimised for a particular observing site. As a consequence turbulence profiling forms a large part of site characterisation studies \citep[see e.g.][]{Vernin2011,Schock2009}. These studies produce many measurements of the profile at a particular site. However, the majority of \ac{AO} simulations used as part of the instrument design process \citep[see e.g.][]{Reeves2016, Conan2014, Rigaut2013, Basden2007}  are Monte-Carlo in nature and require long simulation times and many repeats of the simulation to produce results for a single set of atmospheric conditions. It is therefore not feasible to run simulations on many thousands of turbulence profiles to fully characterise \ac{AO} performance for a particular site. Thus the large dataset of measured turbulence profiles must be reduced to a small set that is in some way representative of the dataset as a whole.

If the turbulence profile at a site were to show very little temporal variation, this task is relatively simple; the average integrated $C_n^2 (h)$ values in each altitude bin for example would give a good approximation of the profile at all times. However for most observing sites the profile varies greatly on timescales from minutes to seasons. In these cases such a method averages out features that are only present in a subset of the data, resulting in a profile that may never have been measured and is therefore not representative of the dataset. An instrument optimised to such a profile would not perform as expected under real world conditions.




Here we put forward a method of obtaining a set of representative turbulence profiles at such a site by employing hierarchical clustering to provide a quantitative classification of profiles. This allows us to separate profiles with different structure and maintain the features in the profile whilst still reducing a large dataset to a small set of profiles.

An example of a site with large variation in the structure of the turbulence profile is ESO Paranal, Chile. A 20 month long campaign using a Stereo-SCIDAR (SCIntillation Detection And Ranging) instrument \citep{Shepherd2014} mounted on one of the \acp{AT} has yielded a set of over ten thousand high resolution (250 m altitude bins) measurements of the turbulence profile at Paranal \citep{Osborn2018a}. We apply the clustering method to this dataset to obtain a small set of turbulence profiles that we validate by comparing distributions of integrated atmospheric parameters. By ensuring the clustered profiles represent the dataset in terms of these parameters we validate them in an atmospheric sense without reference to any particular \ac{AO} system.

We can make the assumption that the free atmosphere turbulence at Paranal is similar to Cerro Armazones, the site of the planned European \ac{ELT}, since they are separated by only around 20 km distance and by around 500 m in altitude. As such this work is relevant to both sites. 

In section \ref{sec:clustering} we present an overview of hierarchical clustering and our method of extracting a small set of turbulence profiles from a large dataset. In section \ref{sec:paranal} we apply this method to the Stereo-SCIDAR dataset from Paranal to obtain a small set of clustered profiles, with comparisons to other turbulence profiles for Paranal. Conclusions are in section \ref{sec:conclusions}.



\begin{figure*}
	\centering
    \includegraphics[]{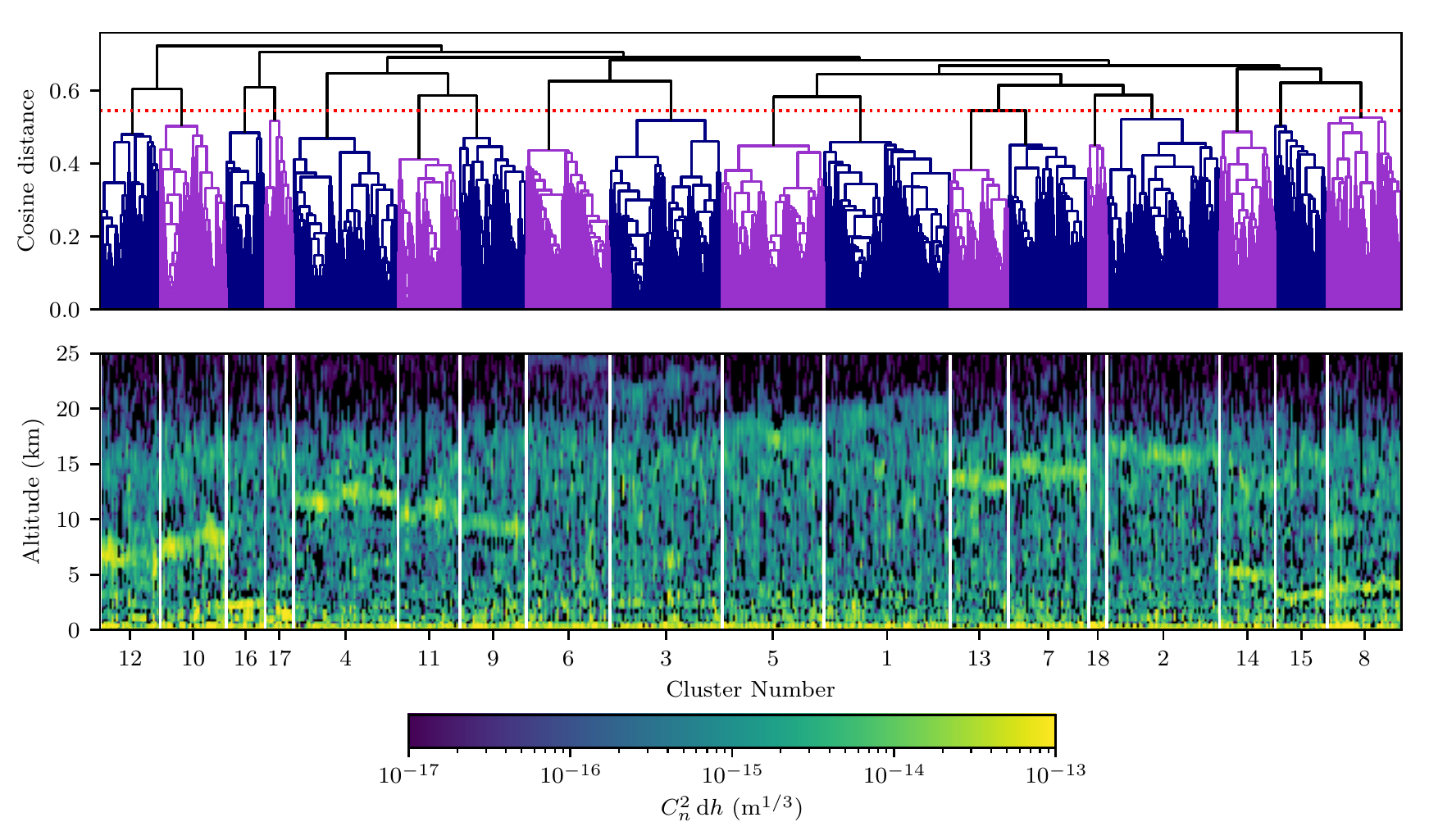}
    \caption{\textit{Upper}: Dendrogram representing average linkage agglomerative hierarchical clustering of the ESO Paranal Stereo-SCIDAR dataset using the cosine distance metric. Branches below a cutoff distance of 0.55 (indicated by the dashed red line) are coloured alternately to indicate 18 clusters. \textit{Lower}: The turbulence profiles in the dataset, ordered according to the leaves of the dendrogram, with the partitioning into 18 clusters indicated by verical white lines. Each cluster is assigned a number according to its size, with 1 being the largest cluster and 18 the smallest.}
    \label{fig:dendrogram}
\end{figure*}

\section{Clustering} \label{sec:clustering}

Cluster analysis allows underlying structure in large datasets to be ascertained by partitioning the data into subsets, known as clusters. There are many different ways to perform clustering on a dataset but here we focus on hierarchical clustering \citep[chapter 4]{Everitt2011}. We settle on this particular variety of clustering for two reasons. Firstly, it allows easy switching and comparison of distance metrics, specifically non-euclidean distance metrics that are particularly effective in this case. Secondly, the clustering can be visualised by the use of a dendrogram (see Fig. \ref{fig:dendrogram}). At the lowest level we have each element in the dataset represented by a vertical line, known as leaves. As we move up the dendrogram to larger distances elements are merged into clusters represented by the joining of two vertical lines into one. To define a certain number of clusters, we cut the dendrogram horizontally at a particular distance and count how many vertical lines (clusters) are intersected. In our case the dendrogram is most useful as a check that the clustering produces sensible results, especially when coupled with the dataset ordered according to the leaves as also displayed in Fig. \ref{fig:dendrogram}.

\subsection{Distance Metrics}

The input to a hierarchical clustering algorithm is the distance matrix $D$. For a dataset of $n$ observations of $p$ variables (in this case $C_n^2\,\mathrm{d}h$ in $p$ altitude bins), $D$ is an $n\times n$ matrix whose components $\delta_{ij}$ represent the pairwise distances between all the observations using a given metric. The choice of the distance metric can have a large impact on the resulting clustering. The most commonly used metric is the euclidean distance:
\begin{equation}
\delta_{ij}^{\mathrm{euc}} = \sqrt{\sum_{k=1}^{p}{(\mathbf{x}_{ik} - \mathbf{x}_{jk})^2}},
\label{eq:euc}
\end{equation}
where $\mathbf{x}_{ik}$ and $\mathbf{x}_{jk}$ represent the $k$th variables in two measurements of the turbulence profile $\mathbf{x}_i$ and $\mathbf{x}_j$ \citep[p. 49]{Everitt2011}. This metric forms the basis of popular clustering algorithms such as K-means \citep{Hartigan1975}. However for profiling data spanning several orders of magnitude in $C_n^2(h)$ the euclidean distance proves to be very sensitive to outliers. As a result, clusters produced using the euclidean distance tend contain a small number of extreme but very similar profiles, while assigning all other profiles (often over half the dataset) to a single large cluster. 

As an alternative, we found the cosine or angular distance to produce favourable results, defined as the normalised dot product
\begin{equation}
\delta_{ij}^{\mathrm{cos}} = 1 - \frac{\mathbf{x}_i \cdot \mathbf{x}_j}{\|\mathbf{x}_i\|_2 \|\mathbf{x}_j\|_2},
\label{eq:cosine}
\end{equation}
where $\|\mathbf{x}\|_2$ denotes the L2 norm of the vector $\mathbf{x}$. For positive data this metric is bound between 0 and 1. The cosine distance is less sensitive to outliers in our case and produces more reasonable clustering for turbulence profiles.

\begin{figure}
\centering
\includegraphics[]{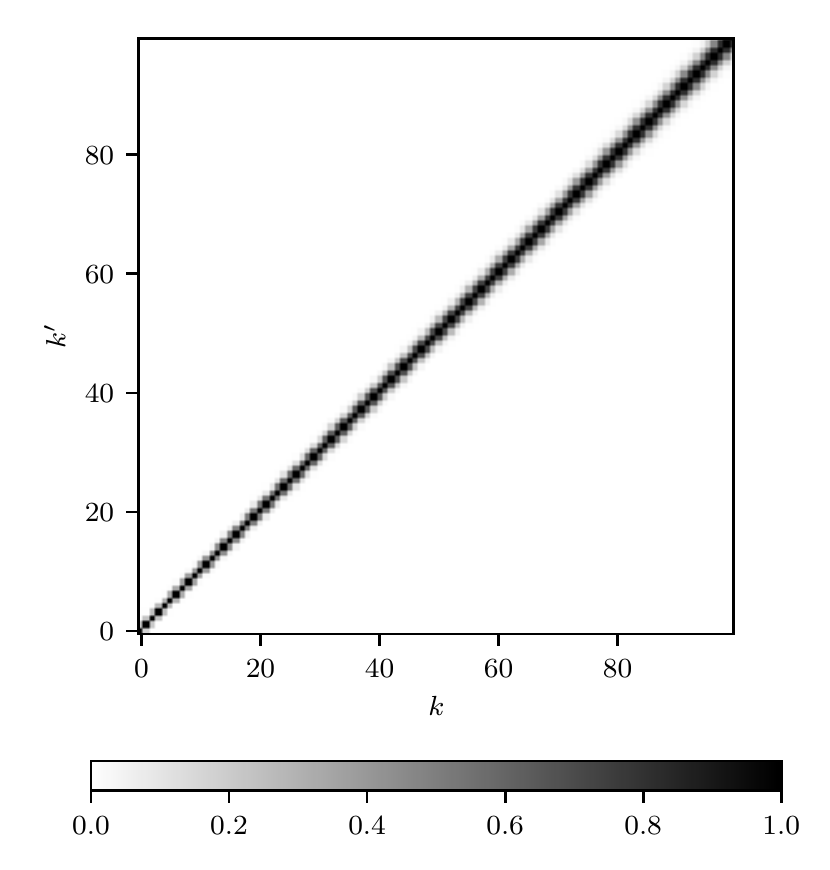}
\caption{\label{fig:S} Similarity matrix $S$ between altitude bins for the Stereo-SCIDAR at Paranal, using an average stellar separation of 12.5", wavelength 500 nm and conjugate altitude $h_{\mathrm{conj}}=-3$ km.}
\end{figure}

In calculating the distance matrix with profile measurement vectors $\mathbf{x}_i$ we have made the implicit assumption that all the components of the vector (altitude bins) are independent. This means that the height of the turbulent layer is not taken into account in the clusters and as such layers that are close in altitude are considered as similar in the distance matrix as layers far apart in altitude. This is not ideal especially since we are dealing with measurements with finite altitude resolution. We therefore modify the cosine metric as described in \citet{Sidorov2014}. By introducing a $p \times p$ matrix $S$ describing the similarity between vector components we obtain the soft cosine distance
\begin{equation}
\delta_{ij}^{\mathrm{softcos}} = 1 - \frac{\sum_k^p\sum_{k^\prime}^{p} S_{k k^\prime}\mathbf{x}_{ik}\mathbf{x}_{jk^\prime}}{\sqrt{\sum_k^p\sum_{k^\prime}^p S_{k k^\prime} \mathbf{x}_{ik} \mathbf{x}_{ik^\prime} } \sqrt{\sum_k^p\sum_{k^\prime}^p S_{k k^\prime} \mathbf{x}_{jk} \mathbf{x}_{jk^\prime}}},
\label{eq:softcosine}
\end{equation}
where both $k$ and $k^\prime$ run through vector components. For $S = \mathds{1}$ this reduces to the cosine distance described in Equation \ref{eq:cosine}. The altitude resolution of the Stereo-SCIDAR is given by 
\begin{equation}
\delta h = 0.5 \frac{\sqrt{\lambda |h-h_{\mathrm{conj}}|}}{\theta},
\label{eq:SCIDARresolution}
\end{equation}
where $\lambda$ is the operating wavelength, taken here to be 500 nm, $h_{\mathrm{conj}}$ is the conjugate altitude of the imaging plane (for the Stereo-SCIDAR at Paranal $h_{\mathrm{conj}}=-3$ km) and $\theta$ is the separation of the double star used to compute the turbulence profile \citep{Avila1997}. We define each row $k$ of $S$ as a gaussian with mean $h_k$ and full width half maximum defined by Equation \ref{eq:SCIDARresolution}. Each row is normalised such that all $S_{kk} = 1$. The widths of these gaussians correspond very well to the response functions of the instrument \citep{Shepherd2014}. The similarity matrix $S$ used for the Stereo-SCIDAR data is shown in Fig. \ref{fig:S}. This process ensures that the distance between profiles as defined by our metric takes into account the finite altitude resolution of the instrument.

\subsection{Clustering process}

The second choice that must be made in hierarchical clustering after the distance metric is the method of defining the inter-cluster distance or linkage. Here we use average linkage, where the inter-cluster distance is defined as the mean pairwise distance between the members of the two clusters. 

A description of the process we employ to perform agglomerative hierarchical clustering is as follows:
\begin{enumerate}
\item{Compute pairwise distance matrix $D$ for the chosen metric.}
\item{Merge the two closest elements.}
\item{Define the new distance from this cluster to the rest of the elements according to the chosen inter-cluster distance.}
\item{Repeat (ii) and (iii) until there are two remaining clusters that are merged into one representing the whole dataset.}
\end{enumerate}

The clustering was performed in python using the hierarchy module in SciPy, which for average linkage clustering utilises the nearest-neighbours chain algorithm \citep[see e.g.][]{Mullner2011}.

\subsection{Data preprocessing}
\label{sec:preprocessing}
The turbulence profiles contain many zero measurements. Usually these occur when turbulence in an altitude bin is below the sensitivity of the instrument but also can be a result of noise in the data post processing pipeline. While it is tempting to treat all zero values as missing data and remove them from the analysis, this can have a profound effect on the calculation of distance between profiles. Thus we choose not to remove these zero measurements before clustering.

\begin{figure}
\centering
\includegraphics[]{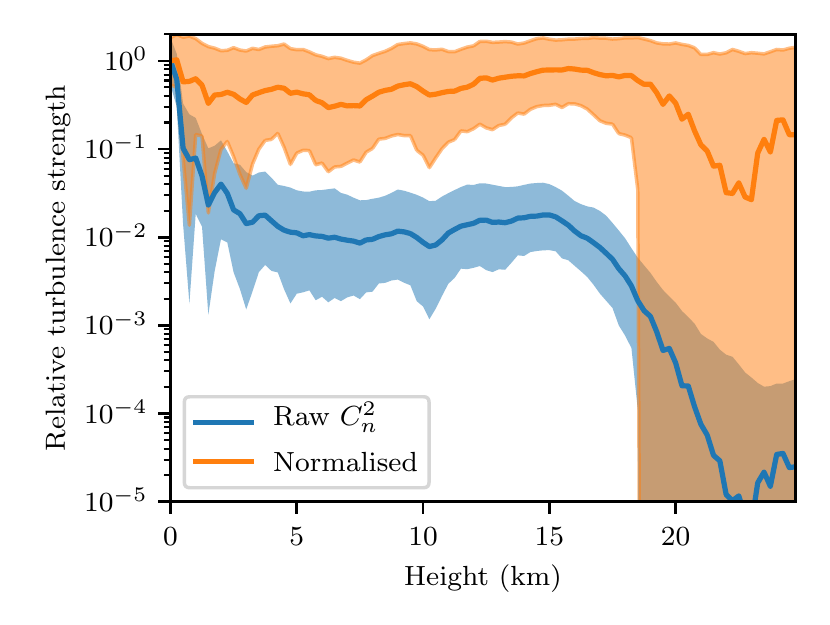}
\caption{\label{fig:norm}The effect on the median (solid line) and interquartile range (shaded areas) of normalisation by dividing each altitude bin by its mean value. Turbulence strength is defined relative to the median value of the first (0 m) bin.}
\end{figure}

\begin{figure}
\centering
\includegraphics[]{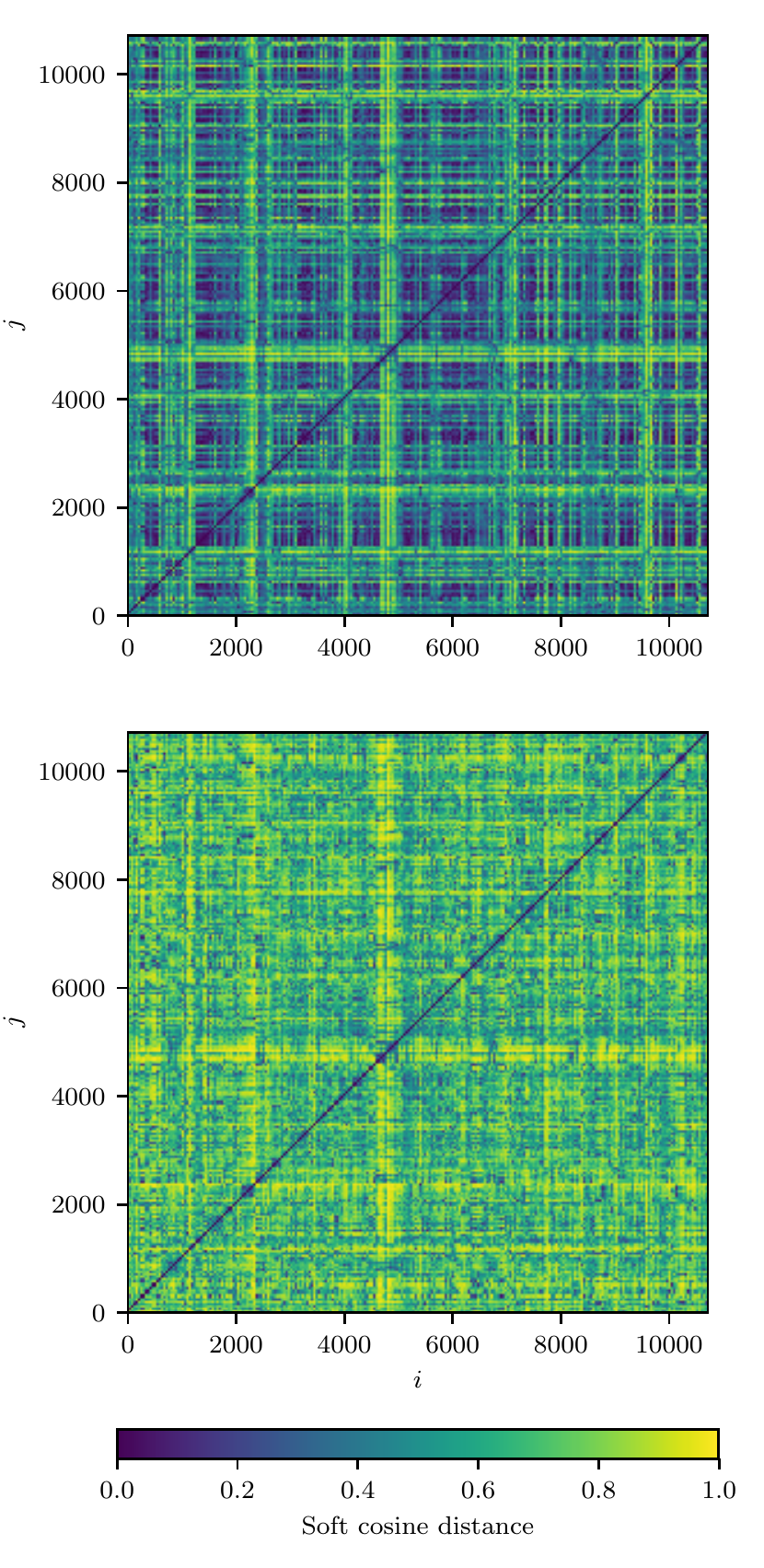}
\caption{Pairwise distance matrices calculated using the cosine metric defined in Equation \ref{eq:cosine} for the Paranal Stereo-SCIDAR dataset of over 10000 turbulence profiles. \textit{Top:} Raw $C_n^2$ measurements. \textit{Bottom:} Profiles normalised by dividing by the mean value in each altitude bin.}
\label{fig:Ds}
\end{figure}

The dynamic range of $C_n^2$ measurements in the data poses a problem in clustering. The distance between profiles tends to be dominated by strong turbulence since these measurements can be up to 100 times stronger than weak or moderate turbulence (see Fig. \ref{fig:norm}). We are more interested in the significance of turbulence, i.e. whether turbulence is high or low relative to the average level of turbulence at a particular height. The $C_n^2$ measurements in each altitude bin are log-normally distributed but the censored nature of the data, where measurements below a sensitivity limit are recorded as zeros, means that we cannot log transform the data and perform the common procedure of subtracting the mean and dividing by the standard deviation for each altitude bin. Instead we find that simply dividing by the mean of each altitude bin is effective in ``flattening'' the profiles, reducing the importance of strong ground layer bins and effectively increasing the importance of weak high layer turbulence such that turbulence at all heights is considered approximately equally in the clustering. The effect of this normalisation on the distance matrix can be seen in Fig. \ref{fig:Ds}. Note that the profiles are additionally L2 normed when the cosine distance is used.

\subsection{Determining the number of clusters}
\label{sec:number}
We seek to cluster turbulence profiles until they are separated according to their structure, such that we can extract a profile from each producing a representative set of profiles. To quantify this we employ two metrics, the within cluster variance and the silhouette score. 

We define the within cluster variance as the sum of the distances of the members of each cluster to the profile we extract as the centre of that cluster. We determine the distance with the same soft cosine metric used in the clustering:
\begin{equation}
W_N = \sum_{m=1}^N \sum_{i=1}^{n_m}{\delta^{\mathrm{softcos}}(X_{im}, X^*_m)}
\label{eq:var_frac}
\end{equation}
where $n_m$ is the number of profiles in cluster $m$, $N$ is the total number of clusters, the $X_{im}$ are all the profiles in cluster $m$ and $X^*_{m}$ is the centre of cluster $m$. The quantity $W_N$ is analogous to the within cluster sum of squares that is minimised in K-means clustering, with the squared euclidean distance substituted for the cosine distance and the cluster centroid $\bar{X}_m$ substituted for our more general cluster centre $X_m^*$. As we increase the number of clusters $N$, $W_N$ will decrease rapidly at first with the gradient falling off as the clustering becomes less effective. It is at this point that we define the number of clusters, a technique known as the elbow method. 


The second metric is the silhouette score \citep[chapter 5]{Kaufman2005}. This metric is defined for a single measurement $i$ as 
\begin{equation}
s_i = \frac{b_i - a_i}{\max{\{a_i,b_i}\}},
\label{eq:silhouette}
\end{equation}
where $a_i$ and $b_i$ are quantities dependent on the distance matrix $D$. $a_i$ represents the average distance between measurement $i$ and all the other members of the cluster $i$ is assigned to. Conversely, $b_i$ represents the average distance between $i$ and all the members of every other cluster. If $a_i > b_i$ resulting in $s_i < 0$ then this profile is on average closer to members of other clusters and is probably assigned to the wrong cluster. If $b_i > a_i$ then $s_i > 0$ and the profile is probably assigned to the correct cluster. A more positive silhouette score is therefore indicative of better clustering. $s_i$ is by definition bounded between -1 < $s_i$ < 1. By taking the mean silhouette score over all members of the dataset $s=\frac{1}{n}\sum_i{s_i}$ we gain insight into the quality of clustering over all clusters.

These two metrics are chosen since, while not completely independent of one another, they incorporate distinct parts of the clustering process. The silhouette score depends solely on pairwise distances between profile measurements described in the distance matrix, whereas the within cluster variance also includes our chosen centre for each cluster $X^*$. This allows us to draw a more robust conclusion as to the number of clusters in the dataset. 

\subsection{Cluster centres}

After performing the clustering and partitioning our dataset we must extract a single turbulence profile from each cluster. The resulting profiles can vary greatly depending on the method used, so we present two methods and hence two sets of turbulence profiles here. 

The simplest way to extract a profile from a cluster is to take an average of each altitude bin in a cluster. More specifically, we take the mean profile in our normed space, then un-normalise this profile and adjust it such that the integrated strength of the profile coincides with the median seeing for the cluster. This results in any features of the clustering common to all profiles in a cluster being retained while features belonging only to a subset of profiles will be averaged out as described earlier. The profiles thus produced will be an unrealistic but conservative description of the variability in profile and will represent the profile in the majority of cases.


Alternatively, we have already defined a metric that describes how well a profile fits into a particular cluster --- the silhouette score. The profile in each cluster with the maximum silhouette score is therefore the best fit profile for that cluster according to our distance metric. In this way we can select an individual turbulence profile as the cluster centre. We therefore select the $N$ profiles from the dataset that represent the centre of each of the $N$ clusters. These profiles will not be ``typical" in the sense that they represent the majority of measurements, but will describe a greater amount of variability which would also be useful for \ac{AO} simulation.

\section{Application to ESO Paranal dataset} \label{sec:paranal}

We use the 2018A Stereo-SCIDAR data release described in \cite{Osborn2018a}. The dataset consists of 10691 turbulence profile measurements taken over 83 nights between April 2016 and January 2018. The profiles have 100 equally spaced altitude bins between the ground and 25 km. 

\begin{figure}
\centering
\includegraphics[width=\columnwidth]{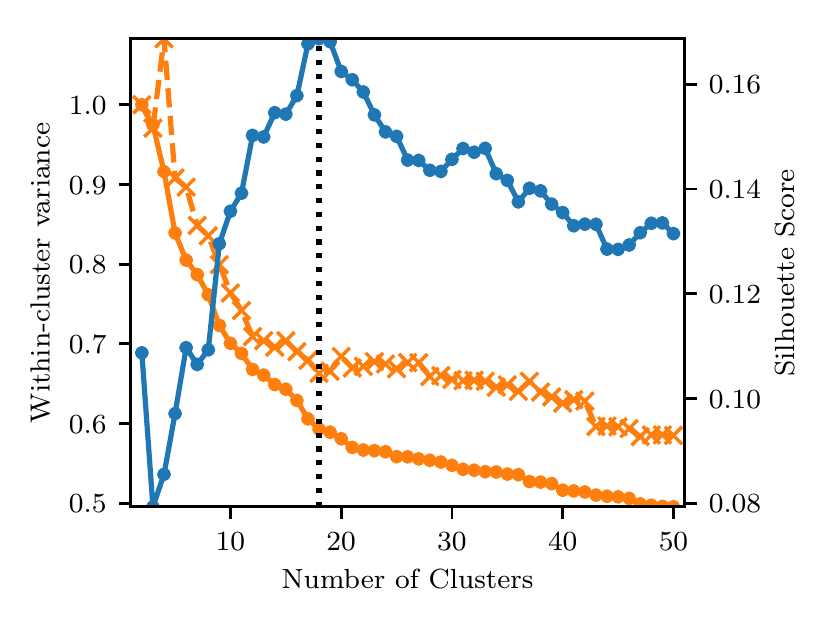}
\caption{Within cluster variance (orange) and silhouette score (blue) for the Paranal Stereo-SCIDAR dataset with increasing numbers of clusters. The two within cluster variance lines represent the two methods of defining the centre of a cluster: average (solid, circular markers) and single profile (dashed, cross markers). Within cluster variance in both cases is normalised to the value at 2 clusters. The dashed vertical line is at 18 clusters.} 
\label{fig:sil_var}
\end{figure}

The metrics for selecting the number of clusters are shown in Fig. \ref{fig:sil_var}. There is a clear peak in the silhouette score at 17--19 clusters. After 19 clusters the silhouette score drops off indicating that further clustering does not improve the quality of the resulting clusters. The within cluster variance in the average centre case shows no clear elbow but a transition from steep to shallow gradient at 15--20 clusters. In the single profile centre case however there is a clearer flattening of the gradient at 18 clusters, corresponding to the centre of the peak in the silhouette score. We therefore choose 18 as our number of clusters.

\begin{figure*}
\centering
\includegraphics[width=\textwidth]{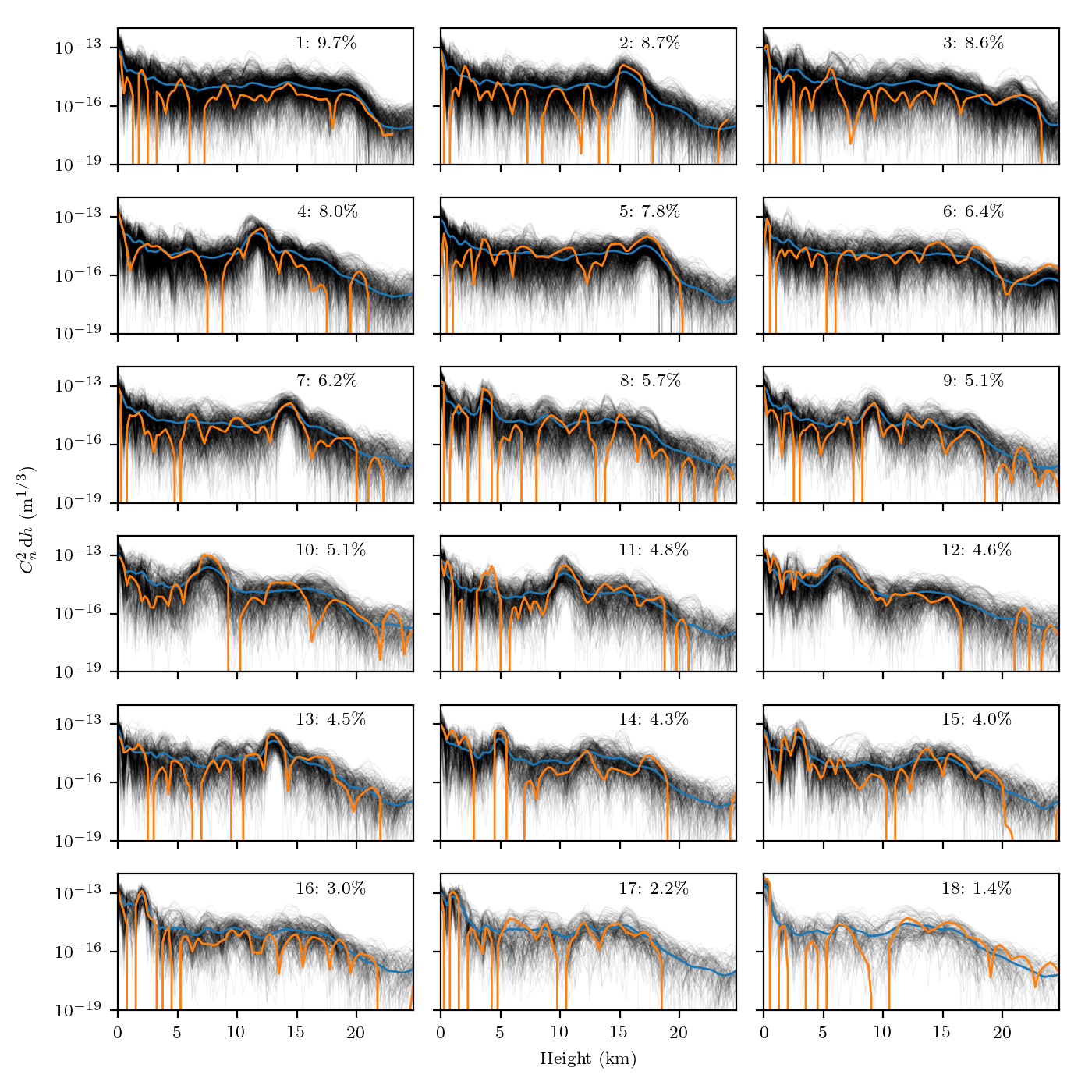}
\caption{The set of 18 full atmosphere turbulence profiles for Paranal extracted through our hierarchical clustering method. Black lines represent every measurement of the turbulence profile in the given cluster. The two methods of obtaining the centre of each cluster are shown as blue (average profile) and orange (single profile) lines. Each cluster is numbered in descending order of the number of profiles in the cluster along with the percentage of all profiles contained in that cluster. Note that these profiles are not normalised.}
\label{fig:all_profiles}
\end{figure*}

The magnitude of the silhouette score is only around 0.17 at the peak which is indicative of structure in the data that has not been captured in the clustering. Indeed we can see from the full set of extracted profiles shown in Fig. \ref{fig:all_profiles} that members of some clusters, especially those containing large numbers of profiles, are fairly inhomogeneous in structure. However, the clustering has for the most part selected and separated profiles with turbulence in strong single layers. This strong single layer is common to almost all profiles in a cluster. The lowest turbulent layers (e.g. clusters 14, 16, 18) tend to be thinner and stronger whereas high layers (e.g. clusters 2, 4, 5) tend to be more spread out and weaker. This may be an instrumental effect due to the reduction in native altitude resolution of the Stereo-SCIDAR with increasing height as described by Equation \ref{eq:SCIDARresolution} and included in the clustering by our use of the soft cosine distance. In total, clusters with significant high altitude ($h \geq 10$ km) layers contain around  55\% of all profiles. We also have separated one ground-layer dominated cluster (18) representing only 1.4\% of profiles. This propensity towards high altitude turbulence is expected from atmospheric parameter statistics for this data: a median isoplanatic angle of 1.75" and fraction of turbulence below 600 m of 0.4 \citep{Osborn2018a}.






\subsection{Comparison Profiles}
\label{sec:comparison}
The most conventional way to reduce a large turbulence profile database to a small set of representative profiles is to first bin the profiles according some integrated parameter, then take an average profile from each bin. The most common parameter used is the integrated strength (seeing), either measured from the profile itself or a contemporaneous measurement from a dedicated seeing monitor such as a DIMM \citep{Sarazin1990}. This is the case for the ESO 35 layer profiles for Paranal \citep{Sarazin2013}, consisting of a profile associated with median seeing and four profiles associated with seeing quartiles. We also produce 18 profiles by binning the Stereo-SCIDAR dataset into 18 seeing bins to provide a more equal comparison to our 18 clustered profiles.


In addition we compare to the good, high and low profiles computed using the method defined in \cite{Sarazin2017}. Rather than binning by the total integrated turbulence strength, the dataset is split into three cases: good seeing, high altitude dominated and low altitude (ground layer) dominated profiles. The average from each of these cases are taken to produce three reference turbulence profiles for Paranal. We also include a profile ``all'' defined as the average of all profiles in the dataset.

\subsection{Validation and comparison}
Whether or not the clustered profiles represent the dataset as a whole is a difficult question to answer since the concept of ``representativeness" can be defined in many different ways. The ultimate aim of this study is to produce a set of turbulence profiles that can be used in \ac{AO} simulation with the knowledge that they reflect the variability in the turbulence profile seen in reality in some meaningful way.

The most direct method of validating the clustered profiles would be using fast analytical \ac{AO} simulation \citep[see e.g.][]{Neichel2008} by comparing relevant \ac{AO} metrics (e.g. tomographic error) over the dataset to the clustered profiles. However, these metrics will depend strongly on the particular system simulated and are therefore beyond the scope of this paper. 

\begin{figure}
\centering
\includegraphics[]{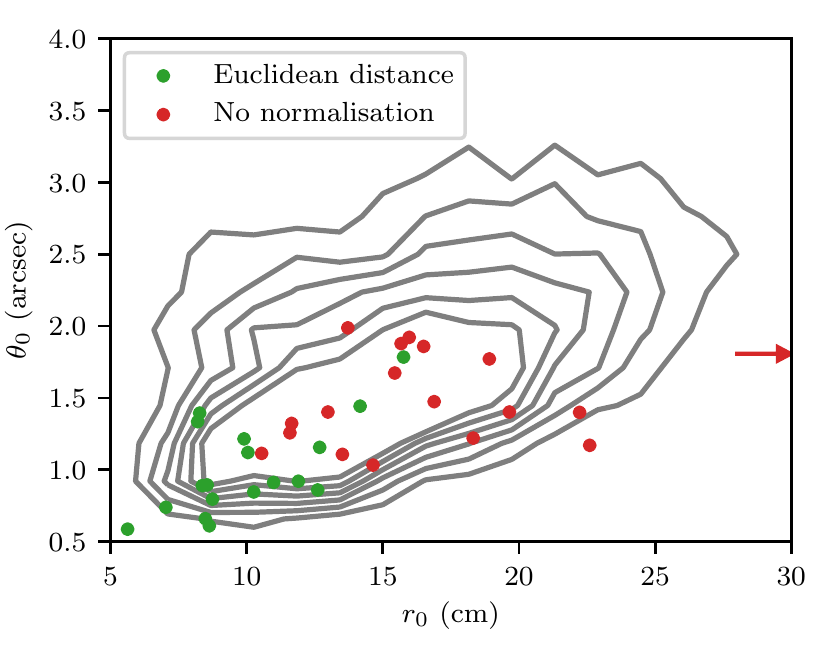}
\includegraphics[]{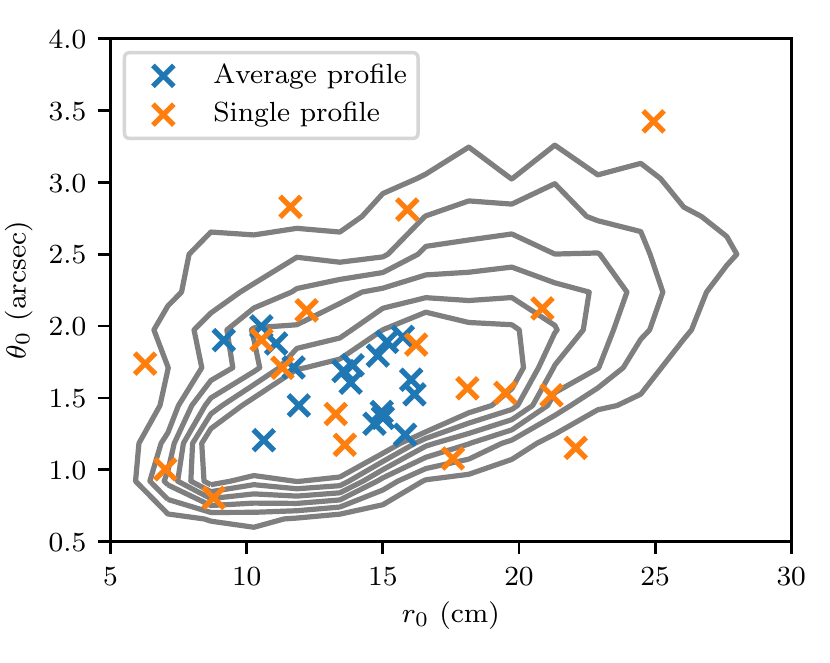}
\includegraphics[]{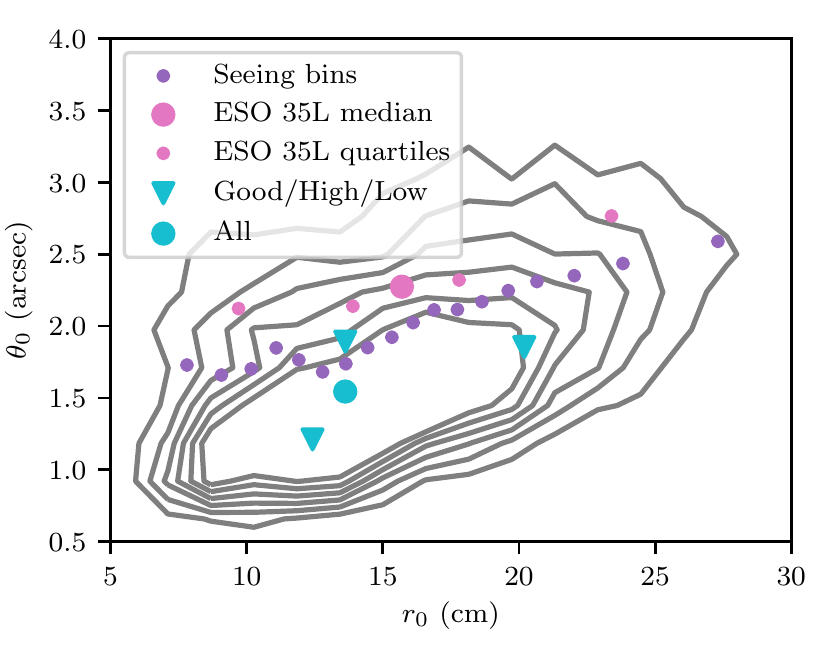}
\caption{\label{fig:params} Distribution of integrated parameters $r_0$ and $\theta_0$ for the entire dataset (contours) and small sets of profiles. \textit{Upper}: Bad clusterings generated through our clustering method with suboptimal parameters. One outlier profile in the no normalisation case with $r_0 = 33$ cm is indicated by an arrow. \textit{Middle}: The two sets of 18 representative clustered profiles with cluster centres defined as average and single profiles. \textit{Lower}: Comparison profiles as discussed in section \ref{sec:comparison}}.
\end{figure}

In the interest of maintaining generality, rather than validating our profiles with \ac{AO} simulation of one or several specific systems, we choose integrated atmospheric parameters as our metrics for validation and comparison to other profiles. While this general atmospheric validation will not necessarily agree with a tomographic \ac{AO} simulation, these parameters serve as reasonable indicators for \ac{AO} performance and are therefore a good compromise given the aforementioned sensitivity of \ac{AO} metrics to the design of the particular system simulated. We choose the Fried parameter $r_0$ \citep{Fried1966} describing the strength of turbulence and isoplanatic angle $\theta_0$ \citep{Roddier1981} describing angular correlation of turbulence, defined respectively as
\begin{equation}
r_0 = \bigg(0.423 k^2 \int_0^{\infty} C_n^2(h) \, \mathrm{d}h\bigg)^{-3/5}
\label{eq:r0},
\end{equation}
\begin{equation}
\theta_0 = \bigg(2.91 k^2 \int_0^{\infty} C_n^2 (h) h^{5/3} \, \mathrm{d}h \bigg)^{-3/5} ,
\label{eq:theta0}
\end{equation}
with $k = 2\pi/\lambda$ the wavevector of light considered (we take $\lambda=500$ nm).  We calculate these parameters for the entire dataset and for our small sets of profiles and the results are shown in Fig. \ref{fig:params}. 

We can see that splitting the dataset into 18 seeing bins and taking an average profile from each produces a set of profiles that by design fits very well with the distribution of $r_0$. However little of the variability in $\theta_0$, a better indicator of the distribution of the turbulence, is described by these profiles. The ESO 35 layer median and quartile profiles behave in the same way. In particular, small values of $\theta_0$ indicating significant high altitude turbulence are poorly represented. The good, high and low profiles provide a better description of the variability $\theta_0$ but are slightly skewed towards larger values of $r_0$ indicating weaker turbulence. The ``all" profile lies in approximately the centre of both distributions as one would expect.

We include in the upper panel of Fig. \ref{fig:params} the distribution of integrated parameters for clustering with some different parameters to those presented above. We find that if we use the euclidean distance instead of the soft cosine distance, the resulting clusters are heavily skewed towards smaller values of both $r_0$ and $\theta_0$. Without normalisation, the clustering produces profiles which better describe the distribution of $\theta_0$ whilst being skewed towards larger values of $r_0$. Combining the soft cosine distance with the normalisation described above (shown in the middle panel of Fig. \ref{fig:params}), we produce profiles that accurately reflect the distributions of both parameters. However, the two methods of defining the centre of a cluster display very different results here. By taking an average profile for each cluster we produce a set of profiles whose integrated parameters are grouped tightly around the centre of the distribution for the dataset. In the case of the $r_0$ distribution this is somewhat by design since we are not sensitive to changes in integrated strength ($r_0$) in our clustering, therefore we produce clusters whose individual distributions of $r_0$ follow approximately the distribution of $r_0$ for the entire dataset. When we set the integrated strength of each of these clustered profiles to the median seeing for that cluster the values will tend to group around the median for the entire dataset. In the distribution of $\theta_0$ however we see a similar tight grouping, with less of the bias towards larger values. 

In contrast, if we take a single profile with the maximum silhouette score as our cluster centre we produce a set of profiles that are spread more widely around parameter space. These profiles therefore describe more extreme variability. Again in the case of $r_0$ this is somewhat by design --- since the clustering is not sensitive to $r_0$ we have essentially randomly sampled the distribution with 18 points, resulting in a wider spread around the parameter space. 

Thus we have produced two sets of profiles that are both representative in different ways. Our average profiles are ``typical'' since they can be used to represent the profile most of the time. The single profiles are not typical since they represent a single measurement at a single time that is unlikely to represent the profile in the majority of times. However, these profiles exhibit more extreme variability in the atmosphere that would be useful in characterising the performance of an \ac{AO} system.

The turbulence profiles presented here are available on request to the author.

\section{Conclusions} \label{sec:conclusions}

We have outlined a method for obtaining a small set of representative turbulence profiles from a large dataset, where all steps of the process are informed by quantitative analysis of the clustering and resulting profiles. 

We applied this method to the Stereo-SCIDAR dataset from ESO Paranal, partitioning over 10,000 measurements into 18 clusters. We have used two methods to obtain the centre of each cluster resulting in two sets of 18 high resolution full atmosphere turbulence profiles with 100 altitude bins between 0 m and 25 km. While the clustering has not preserved all the structural variation in the turbulence profile at Paranal, each cluster is dominated by a single strong turbulent layer, the height of which varies over the full range of altitudes. 

Through analysis of integrated turbulence parameters it has been shown that the two sets of profiles are two distinct forms of ``representative'' profile. Taking the average profile for each cluster results in typical profiles grouped around the centre of parameter space and representing the profile in the majority of cases. Conversely defining a single profile as the cluster centre produces a set of profiles that represent more extreme variability in the dataset. Validation of these profiles for specific instruments using tomographic \ac{AO} simulation remains for a future publication. Additionally, it would be possible to produce a set of profiles representative of the variability in profile for a particular instrument by performing the clustering on \ac{AO} metrics relevant to that instrument (e.g. tomographic error).

Future work will focus on the temporal statistics of these clustered profiles, on both short timescales of minutes to hours and longer seasonal timescales. Analysis of seasonal variability in particular will require more data from the Stereo-SCIDAR to ensure statistically significant results. 


More generally in the context of site characterisation and monitoring, clustering methods can be applied not only to large databases of turbulence profiles but to any multivariate data (e.g. wind, humidity, temperature) in order to extract small sets of representative conditions. Data from existing instruments such as \ac{AO} telemetry or point spread functions could also be used either as input to the cluster analysis or as validation for representative atmospheric conditions. 

\section*{Acknowledgements}

This work was supported by the Science and Technology Funding Council (UK) (ST/P000541/1). OJDF acknowledges the support of STFC (ST/N50404X/1).

Horizon 2020: This project has received funding from the European Union's Horizon 2020 research and innovation programme under grant agreement No 730890. This material reflects only the authors views and the Commission is not liable for any use that may be made of the information contained therein.

PJ acknowledges the support of the National Natural Science Foundation of China (NSFC) (11503018, U1631133) as well as the China Scholarship Council (CSC). 

This research made use of Python including NumPy and SciPy \citep{vanderWalt2011}, Matplotlib \citep{Hunter2007} and Astropy, a community-developed core Python package for Astronomy \citep{Robitaille2013}. We also made use of the Python AO utility library AOtools (\url{https://github.com/AOtools/aotools}).



\bibliographystyle{mnras}
\bibliography{library} 







\bsp	
\label{lastpage}
\end{document}